\documentclass[journal]{IEEEtran}
\pdfoutput=1
\usepackage{graphicx}
\usepackage{subfig}
\usepackage{amsmath,array}
\usepackage{amsfonts}
\usepackage{algorithm,algorithmic}
\usepackage{booktabs,multirow}
\usepackage{enumerate}
\usepackage{url}
\usepackage{hyperref}
\usepackage{color,soul}

\begin{document}

\title{Contrastive Multi-Level Graph Neural Networks for Session-based Recommendation}

\author{Fuyun Wang, 
        Xingyu~Gao$^*$,~\IEEEmembership{Member,~IEEE,}
        Zhenyu~Chen,~\IEEEmembership{Member,~IEEE,}
        Lei Lyu
\thanks{F. Wang and L. Lyu are with School of Information Science and Engineering, Shandong Normal University, Jinan 250358, China.}
\thanks{$^*$Corresponding author: Xingyu Gao is with Institute of Computing Technology, Chinese Academy of Sciences, Beijing, China.}
\thanks{Zhenyu Chen is with Institute of Computing Technology, Chinese Academy of Sciences, Beijing, China.}
}


\markboth{IEEE Transactions on Multimedia,~Vol.~XX, No.~X, November~2022}%
{F. Wang \MakeLowercase{\textit{et al.}}: Contrastive Multi-Level Graph Neural Networks for Session-based Recommendation}

\maketitle

\begin{abstract}
  Session-based recommendation (SBR) aims to predict the next item at a certain time point based on anonymous user behavior sequences. Existing methods typically model session representation based on simple item transition information.  However, since session-based data consists of limited users' short-term interactions, modeling session representation by capturing fixed item transition information from a single dimension suffers from data sparsity. In this paper, we propose a novel contrastive multi-level graph neural networks (CM-GNN) to better exploit complex and high-order item transition information. Specifically, CM-GNN applies local-level graph convolutional network (L-GCN) and global-level network (G-GCN) on the current session and all the sessions respectively, to effectively capture pairwise relations over all the sessions by aggregation strategy. Meanwhile, CM-GNN applies hyper-level graph convolutional network (H-GCN) to capture high-order information among all the item transitions. CM-GNN further introduces an attention-based fusion module to learn pairwise relation-based session representation by fusing the item representations generated by L-GCN and G-GCN. CM-GNN averages the item representations obtained by H-GCN to obtain high-order relation-based session representation. Moreover, to convert the high-order item transition information into the pairwise relation-based session representation, CM-GNN maximizes the mutual information between the representations derived from the fusion module and the average pool layer by contrastive learning paradigm. We conduct extensive experiments on multiple widely used benchmark datasets to validate the efficacy of the proposed method. The encouraging results demonstrate that our proposed method outperforms the state-of-the-art SBR techniques.
\end{abstract}


%
\IEEEpeerreviewmaketitle

\section{Introduction}
%
%
%
%
\IEEEPARstart{A}{s} an important way to alleviate information overload, recommendation systems~\cite{min2019food,gao2019hierarchical,chen2020learning,quintanilla2020adversarial,sang2020context} play an important role in many e-commerce platforms, online entertainment platforms, and web search applications. Typically, on many websites, a user engages in brief interactions in a scenario where the system is not logged on. These interactions between the anonymous user and the system are organized into a session. The task of session-based recommendation (SBR) is to predict the user's next interaction based on the user's historical interaction session. Compared with conventional sequential recommendations methods, SBR methods model user preferences by capturing item transitions in the sessions and have attracted widespread attention in recent years \cite{li2017neural}, \cite{liu2018stamp}, \cite{shani2005mdp}.

Given the characteristics of session-based data, many effective recommendation paradigms are proposed in early studies (e.g., item co-occurrence relations-based recommendation methods \cite{sarwar2001item} and Markov chain-based recommendation methods \cite{rendle2010factorizing},  \cite{he2016fusing}). However, these approaches either ignore users' sequential behavior patterns or fail to capture complex item transitions in sessions. With the rapid growth of Recurrent Neural Networks (RNNs), many RNNs-based recommendation methods have achieved encouraging results in the fields of SBR \cite{hidasi2015session}, \cite{yu2016dynamic}, \cite{wu2017recurrent}, \cite{li2017neural}, \cite{jannach2017recurrent}, \cite{tan2016improved}, \cite{hidasi2018recurrent}. Despite the promising progress, RNNs-based approaches overemphasize the sequential patterns of item transitions and thus suffer from overfitting problems. Besides, since RNNs-based approaches only model the relations between adjacent items and fail to capture implicit connections among all items, these methods still have shortcomings in their ability to mine the features of session-based data. Recently, self-attention-based approaches (e.g. SASRec) \cite{subakan2021attention}, \cite{kang2018self}, \cite{li2020time}, \cite{ren2020sequential}, \cite{wang2019dmran}, \cite{sun2019bert4rec} have been proposed to capture long-term dependencies in the sequence. But these kinds of approaches have learned too much item transition information which is not related to the current session and thus increases computational complexity.

In recent years, graph neural networks (GNNs) have been widely used in SBR \cite{pan2020star}, \cite{wu2019session}, \cite{qiu2019rethinking}, \cite{xu2019graph}, \cite{chen2020handling}. SR-GNN \cite{wu2019session} models complex transitions among items by adopting a gated graph neural network. But SR-GNN only calculates the relative importance of each item to the last item when learning session representation, which fails to capture the specific item transition pattern within the session. Compared with SR-GNN, FGNN \cite{qiu2019rethinking} fully considers the inherent order of the item transition patterns and has achieved better performance. Although these methods alleviate the problem of previous work failing to capture implicit connections among all items to some extent, they suffer from two inherent algorithmic drawbacks. On the one hand, existing methods only use the current session to make recommendations, ignoring item transition patterns in other related sessions. On the other hand, existing methods model each session as a directed subgraph and regard item transitions as pairwise relations, failing to exploit the high-order information of item transitions.

To tackle the above issues, we propose a novel contrastive multi-level graph neural network (CM-GNN) to model the complex item transition patterns over all the sessions. Specifically, we design a local-level graph convolutional network (L-GCN) and a global-level graph convolutional network (G-GCN) to capture pairwise relations of the current session and all the sessions, respectively. To be specific, G-GCN can capture item transition patterns related to the current session from other sessions, which is helpful to learn more accurate item representations. To learn pairwise relation-based session representation, an attention-based fusion module is designed to fuse the item representations outputted by L-GCN and G-GCN. Furthermore, CM-GNN designs a hyper-level  graph convolutional network (H-GCN) for capturing high-order relations. By averaging the item representations generated by H-GCN, we obtain high-order relation-based session representation. Since the above two levels of session representation can be seen as two different views which describe the same session, CM-GNN converts the high-order item transition information into the pairwise relation-based session representation by maximizing the mutual information between the different session representations through contrastive learning. The main contributions of our work can be concluded as follows:

\begin{itemize}
\item[$\bullet$]  We propose a novel contrastive multi-level graph neural network (CM-GNN) for session-based recommendation (SBR), which can model complex and high-order item transition patterns over all the sessions.
\item[$\bullet$]  We design local-level graph convolutional network (L-GCN), global-level graph convolutional network (G-GCN) and hyper-level graph convolutional network (H-GCN) for capturing pairwise relations and high-order relations.
\item[$\bullet$]  We obtain a pairwise relation-based session representation through an attention-based fusion module and a high-order relation-based session representation through an average pool layer, respectively, and incorporate high-order item transition information with pairwise item transition information through contrastive learning.
\item[$\bullet$]  We conduct comprehensive experiments on three benchmark datasets. The encouraging results show that our proposed CM-GNN outperforms the state-of-the-art methods.
\end{itemize}

The remainder of this paper is organized as follows. Section \ref{sec2} reviews the related work. Section \ref{meth} describes the proposed CM-GNN in detail. Section \ref{expe} presents extensive experiments. Finally, Section \ref{conclu} concludes the paper.

\section{Related work}\label{sec2}
\subsection{Session-based Recommendation}
The task of SBR is to predict users' next interaction according to their historical behavior sequences. Conventional SBR methods include item co-occurrence relation-based methods \cite{sarwar2001item} and Markov chain-based methods \cite{rendle2010factorizing} \cite{he2016fusing}, \cite{zimdars2013using}. Item co-occurrence relation-based methods are recommended based on the similarity between items and these methods fail to capture sequential patterns within sessions. Markov chain-based approaches use latest users' click to model users' short-term preferences, ignoring useful information used to model users' long-term preferences.

With the rapid development of neural networks in the field of natural language processing, deep learning-based methods have also achieved advanced performance in SBR. Hidasi et al. \cite{hidasi2015session} first apply Recurrent Neural Networks (RNNs) to sequential recommendation and propose a GRU4Rec model. Afterward, Li et al. \cite{li2017neural} introduce attention mechanisms into GRU4Rec and capture more representative item transition information. Liu et al. \cite{liu2018stamp} propose a short-term memory network named STAMP that uses attention mechanisms to replace RNNs encoding, which captures the users' general interests from the long-term memory of session context and takes into account the user's current interest from the short-term memory of the last click. SASRec \cite{kang2018self} proposed by Wang et al. is another classical method in sequential recommendation, which models the users' historical behaviors through self-attention mechanism and fully captures the transition information of the item. Since Convolutional Neural Networks (CNNs) do not depend on the previous state and are sufficiently parallel in calculation, Wang et at. \cite{wang2019collaborative} utilize memory networks to predict users' intent based on collaboration sessions. The disadvantages of RNNs-based models and CNNs-based models are that they only consider the relations of adjacent items and fail to capture the complex relations among non-adjacent items \cite{wang2019sequential}, \cite{wang2021survey}.

In recent years, Graph Neural Networks (GNNs) have achieved promising results in capturing complex relations between nodes \cite{wu2020comprehensive}, \cite{kipf2016semi}, \cite{he2020lightgcn}, \cite{pan2020star}, \cite{chen2021hybrid}. In the field of SBR, Wu et al. \cite{wu2019session} first propose the SR-GNN which used the graph neural network for sequence modeling. SR-GNN regards the problem of sequence modeling as a problem of graph modeling and learns the hidden vector representation of each item through a gating graph neural network. With the successful application of SR-GNN, some variants of GNNs have also achieved encouraging performance. GC-SAN \cite{xu2019graph} uses gating graph neural network (GGNN) to learn transition information between items and to learn long-term interests of users based on multiple self-attention layers. By studying inherent order of the item transition pattern in the sequence, FGNN \cite{qiu2019rethinking} compensates for the SR-GNN's failure to take into account the transition pattern of specific items within the session, further enhancing the recommendation performance.

Although the above approaches have achieved promising performance in SBR, they still fail to obtain more useful information. First of all, the above methods construct each session into a directed subgraph that regards item transitions as pairwise relations. However, in the real-world scenario, there are often high-order relations among items, which can not be captured by the above methods. In addition, since the above methods are only for single-session graphs and do not take cross-session information into account, we often obtain the inaccurate session representations. Wang et al. \cite{wang2020next} propose HyperRec which is the earliest work used by hypergraph conventional network for sequential recommendation. HyperRec models the short-term preferences of users by constructing the sequence as a hypergraph \cite{yadati2018hypergcn}, \cite{feng2019hypergraph}. Afterward, Xia et al. \cite{xia2020self} propose S$^2\raisebox{0mm}{-}$DHCN which is a dual-channel hypergraph convolutional network for SBR. S$^2\raisebox{0mm}{-}$DHCN models users' short-term interests based on hypergraph channels and constructs self-supervision signals based on line graph channels to enhance hypergraph modeling. In order to capture relevant item transition information from other sessions, CA-TCN \cite{ye2020cross} and GCE-GNN \cite{wang2020global} first learn the item representation based on the session graph and then calculate the local-level and global-level session representations respectively. Unlike CA-TCN, GCE-GNN constructs a global-level graph based on all the sessions and combines local-level session representations representing the user's short-term interests with global-level session representations representing the user's long-term interests to obtain the final session representation. GCE-GNN is the state-of-the-art method and has achieved the best performance in SBR. However, GCE-GNN only takes into account pairwise relations and fails to model the high-order relations. In addition, existing studies fail to combine pairwise relations with high-order relations, which is made up by our work.

\subsection{Contrastive Learning Recommendation}
Recently, contrastive learning \cite{chen2020simple}, \cite{hassani2020contrastive}, \cite{oord2018representation}, \cite{hjelm2018learning}, \cite{trinh2019selfie}, \cite{wuw2021self} has been widely applied to mine its own supervision information from large-scale unsupervised data by using pretext tasks. It is first applied to the field of computer vision \cite{sermanet2018time}, \cite{bachman2019learning}, \cite{ji2019invariant}, \cite{ye2019unsupervised}, \cite{he2020momentum},  and then make important progress in the field of audio processing \cite{ma2020active}, and natural language processing \cite{devlin2018bert}, \cite{arora2019theoretical}, \cite{fang2020cert}, \cite{iter2020pretraining}. An important branch of contrastive learning is mutual information maximization \cite{hjelm2018learning}, \cite{xin2020self}, which maximizes the mutual information between these views by using views from the same input as positive samples and views from different inputs as negative samples. S$^3\raisebox{0mm}{-}$Rec \cite{zhou2020s3} is a classical application of contrastive learning in sequential recommendation, which maximizes the mutual information between attribute, item, and sequence views by masking different levels of granularity of the contextual information. In the field of graph representation learning, contrastive learning has also made encouraging progress. SGL \cite{wu2021self} aims to explore contrastive learning on the user-item graph. By changing the structure of the graph to generate multiple views of a node, SGL maximizes the agreement between different views of the same node and the views of other nodes, enriching the feature representation of the node. In the field of SBR, S$^2\raisebox{0mm}{-}$DHCN models session-based data as a hypergraph, and then constructs self-supervision signals based on the hypergraph-induced line graph, enhancing the modeling of the hypergraph by maximizing the mutual information between session representations learned by different graphs. Recently, contrastive learning has also been widely used in the field of social recommendation \cite{yu2021socially}, and GroupIM \cite{sankar2020groupim} and MHCN \cite{yu2021self} have gained more comprehensive user representation by constructing self-supervision signals from users' social relationships. Unlike the above approaches, since the session typically consists of brief interactions by anonymous users, this work focuses on exploring contrastive learning on item transitions in SBR to improve user intent learning.

\section{Methodology}\label{meth}
In this section, we describe the algorithm details of contrastive multi-level graph neural network (CM-GNN). In subsection \ref{pd}, we describe the problem definitions for session-based recommendation (SBR) and introduce the notion definitions used throughout the paper. In subsection \ref{ov}, we present an overview of CM-GNN. In subsection \ref{gcc}, we describe the local-level graph convolutional network (L-GCN), global-level graph conventional network (G-GCN), and hyper-level graph convolutional network (H-GCN). In subsection \ref{sess}, we describe how to learn the pairwise relation-based session representation (pairwise relation-based session representation is generated by items that contain pairwise item transition information) and the high-order relation-based session representation (high-order relation-based session representation is generated by items that contain high-order item transition information) based on the item representations generated by different GCNs. In subsection \ref{predic}, we describe the prediction process of our model. In subsection \ref{opti}, we show the optimization process of CM-GNN. In the subsection \ref{contras}, we describe how to convert the high-order item transition information into the pairwise relation-based session representation through contrastive learning paradigm.

\subsection{Problem Definition}\label{pd}
Let $ V = \{\upsilon_1, \upsilon_2, ..., \upsilon_n \} $  denote the set of all items, where $ {n} $ is the number of all unique items. Each anonymous session $ {s} $ can be represented by list $ {s} = \{ \upsilon_{s,1}, \upsilon_{s,2}, \upsilon_{s,3}, ..., \upsilon_{s,m} \}  $, where $ {m} $ is the length of $ {s} $ and $v_{s,k} \in V( 1 \leq  {k} \leq {m} )  $  represents an interacted item of an anonymous user within the session $ {s} $. We describe each item $ \upsilon_i \in V  $ with embedding vector $ \textbf{x}^{(t)}_i \in \mathbb{R}^{d^{(t)}} $, where $ d^{(t)} $ is the dimension of item in the $ {t} $-th layer of a GCN. Given a specific session $ {s} $,  the goal of session-based recommendation (SBR) is to predict the next item $ v_{s,m+1} $. Formally, our model aims to output the click probability of all items $ \hat{y} = \{ \hat{y}_1, \hat{y}_1, \hat{y}_3, ..., \hat{y}_n \} $, where $ \hat{y}_i \in \hat{y} $ indicates the likelihood score of clicking item $ {v}_i $. We rank the prediction scores in descending order, and recommend the candidate items ranked in the top-K. In this paper, we adopt bold capital letters to denote matrices, bold lowercase letters to denote vectors, and italic lowercase letters to represent scalars.

\begin{figure*}
  \centering
  \includegraphics [scale=0.7]{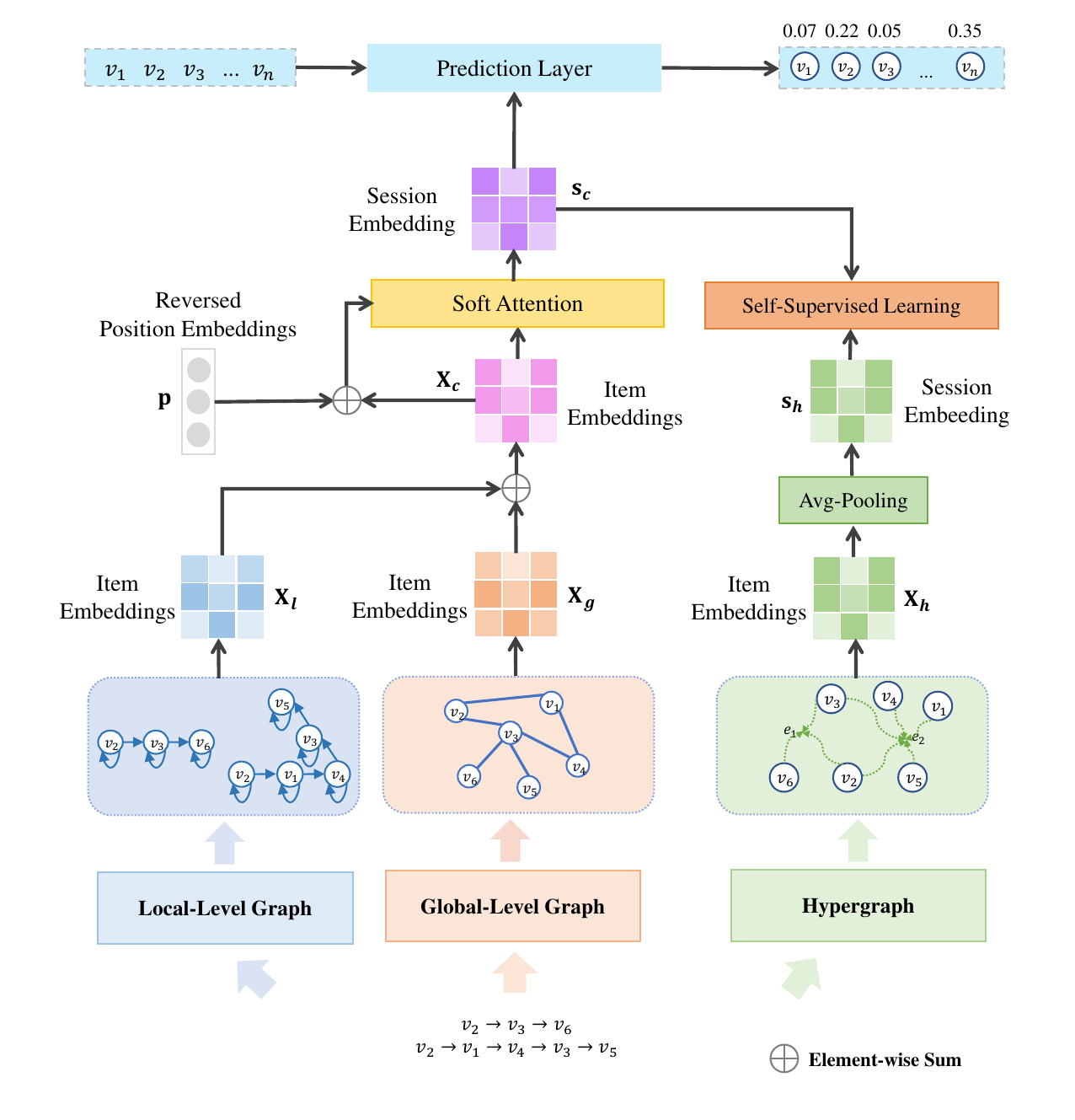}

  \centering  \caption  {The overall framework of our proposed CM-GNN.}
  \label{framework}
\end{figure*}
\subsection{Overview}\label{ov}
We illustrate the framework of CM-GNN in Fig. \ref{framework}. Firstly, we build all the sessions, the current session, and all the sessions as a global-level graph (G-G), local-level graph (L-G), and hyper-level graph (H-G) respectively. Then for each type of graph, we apply global-level graph convolutional network (G-GCN), local-level graph convolutional network (L-GCN), and hyper-level graph convolutional network (H-GCN) to capture global-level pairwise relations, local-level pairwise relations, and high-order relations respectively. Afterward, we use an attention-based fusion module to learn the pairwise relation-based session representation by fusing the global-level item representations and the local-level item representations. We also obtain the high-order relation-based session representation by averaging the hyper-level item representations. Moreover, we integrate contrastive learning into CM-GNN and incorporate high-order item transition information into the pairwise relation-based session representation by maximizing the mutual information across different levels of session representation. Finally, we obtain the next-click one for a given session by computing the scores of each candidate item.

\subsection{Different Levels of Graph Convolutional Network}\label{gcc}
In this section, we introduce the construction process of different levels of graph and describe the corresponding graph convolutional network on them.

\subsubsection{L-GCN}
\begin{figure}
  \centering
  \includegraphics [scale=0.44]{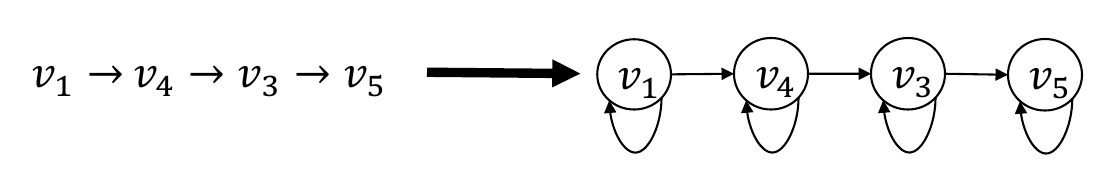}

  \caption  {The construction process of the local-level graph.}
  \label{localgraph}
\end{figure}

To capture the pairwise relations among items in the current session, we construct a L-G for each session $ {s} $. Let $ G_{l} = (\mathcal{V} _{l}, \mathcal{E} _{l}) $ denote L-G, where $ \mathcal{V}_{l} \subseteq V $ and $ \mathcal{E}_{l} = \{e_{ij}^l\mid (v_i,v_j)\mid v_i, v_j \in V \} $ represent the node set and the edge set, respectively. Each edge $ (v_i, v_j) \in \mathcal{E}_{l} $ implies that a user clicks item $ v_j $ after $ v_i $ in the session $ s $ and we capture pairwise item transition patterns of current session based on these edges. To get more item transition patterns, we add self-loop to each item. Moreover, since L-G is a directed graph, the types of directed edge are distinguished into four types based on the item transition order, i.e., $ r_{in}, r_{out}, r_{in-out} $ and $ r_{self} $. Given a specific edge $ (v_i, v_j) $, $ r_{in} $ indicates there is only one transition from $ v_i $ to $ v_j $ while $ r_{out} $ indicates there is only one transition from $ v_j $ to $ v_i $, $ r_{in-out} $ indicates there are both transitions from $ v_i $ to $ v_j $ and from $ v_j $ to $ v_i $ and $ r_{self} $ indicates an item has a self-transition. The construction process of L-G is illustrated in Fig. \ref{localgraph}.

Since not all neighbors of item $ v_i $ have the same importance to item representation learning, we introduce an attention mechanism to measure the contribution of neighbors by learning weights of edges between different nodes. Formally, the attention coefficients are calculated as follows:

\begin{equation}
  \centering
  a_{ij} = LeakyRelu(\textbf{w}^T_{r_{i,j}}(\textbf{x}_{v_i}\odot \textbf{x}_{v_j})),
\end{equation}
where $ a_{ij} $ measures the importance of node $ v_j $ to node $ v_i $ and $ LeakyRelu(\cdot ) $ is chosen as activation function. $ r_{ij} $ is the relation between node $ v_i $ and node $ v_j $. $ \odot $ is the element-wise multiplication operation and $ \textbf{w} _\ast \in  \mathbb{R}^d $ are weight vectors.

There may be more than one neighbor for a same node and the contributions of these neighbors are different when learning node representation. As a consequence, if the representation of the current node is learned based on an equal contribution of its neighbors, it would result in inaccurate learning of node representation. To avoid this problem, we first normalize the attention coefficients through the softmax function:

\begin{equation}
  \centering
  \alpha _{ij} = \frac{\exp (LeakyRelu(\textbf{w}^{T}_{r_{ij}}(\textbf{x}_{v_{i}} \odot \textbf{x}_{v_{j}})))}
  {\sum _{v_{k}\in \mathcal{N} ^{l}_{v_{i}}}\exp (LeakyRelu(\textbf{w}^{T}_{r_{ik}}(\textbf{x}_{v_{i}} \odot \textbf{x}_{v_{k}})))},
\end{equation}
where $ \alpha _{ij} $ denotes the attention coefficients after normalization, $ \mathcal{N} ^{l}_{v_i} $ represents the collection of nodes which are adjacenct to node $ v_i $ in L-G.

Then we obtain output features for each node by calculating the linear combination of the attention coefficient and its corresponding item representation.
\begin{equation}
  \centering
  \textbf{x}^l_{v_i} = \sum \limits_{ v_j \in  \mathcal{N} ^{s}_{v_{i}}} \alpha _{ij} \textbf{x}_{v_j}.
\end{equation}
We use an attention mechanism to aggregate the features of the item itself and its neighbors in the current session and obtain the local-level item representation.

We denote the pairwise local-level item representations as $\textbf{X}_l$, where each row in $\textbf{X}_l$ represents the local-level item representation after L-GCN learning, i.e. $\textbf{x}^l_{v_i}$.

\subsubsection{G-GCN}
\begin{figure}
  \centering
  \includegraphics [scale=0.84]{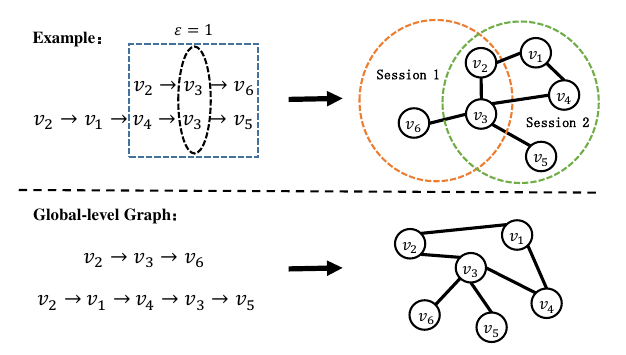}
  \caption  {The construction process of the global-level graph.}
  \label{globalgraph}
\end{figure}

Since a session is generally a short sequence, modeling the current session only based on L-G often suffers from the problem of data sparsity and results in difficulty generating a comprehensive and accurate item representation. To overcome the above problems, we construct G-G over all the sessions to capture the item transition patterns within all sessions. Based on the fact that the same item appears in both the current session and the other sessions, we follow the concept of $ \varepsilon  $-neighbor proposed by GCE-GNN \cite{wang2020global} (i.e., $ \varepsilon  $-neighbor set) to construct G-G. Let $ G_g=(\mathcal{V} _g, \mathcal{E} _g) $ denote G-G, where $ \mathcal{V} _g $ is the node set which contains all the items in $ V $, $ \mathcal{E} _g = \{ e_{ij}^g \mid  (v_i, v_j) \mid v_i \in V, v_j \in \mathcal{N} _{\varepsilon (v_i)} \} $ indicates the edge set, where $ \mathcal{N} _{\varepsilon (v_i)} $ represents the $ \varepsilon  $-neighbor set of item $v_i$. Each edge $ e_{ij} $ corresponds to a pairwise relation between item $ v_i $ and item $ v_j $ existing in the $ \varepsilon  $-neighbor of item $ v_i $ within all sessions. Different from L-G, to distinguish the importance of different neighbors to item $  v_i $, each edge is assigned a weight based on the occurrence frequency of corresponding item transition patterns within all sessions. G-G is an undirected graph and the construction process is illustrated in Fig.  \ref{globalgraph}.

Based on  G-G, we develop G-GCN to capture the pairwise relations derived from all the sessions. Specifically, we first identify the multiple sessions where item $ v_i $ is involved and then capture the relevant item transition patterns. Since not all of the items in $ \mathcal{N} _{\varepsilon (v_i)} $  are relevant to the user preference of the current session, we introduce a session-aware attention mechanism to learn the importance of each neighbor to $ v_i $.

First, to obtain relevance between items in $ \mathcal{N} _{\varepsilon (v_i)} $ and the current session, we calculate representation of the current session by averaging the item representations:

\begin{equation}
  \centering
  \textbf{s} = \frac{1}{m} \sum \limits_{ v_i \in s}\textbf{x}_{v_i},
\end{equation}
where $ s $ denotes the current session and $ m $ is the length of $ s $.

Then, we calculate importance weight of different neighbors to the current session:

\begin{equation}
  \centering
  b_{ij} = \textbf{q}_1LeakyRelu(\textbf{W}_1[(\textbf{s} \odot \textbf{x}_{v_j})\parallel w_{ij}]),
\end{equation}
where $ b_{ij} $ denotes the weight coefficient of different neighbors of node $ v_i $ in G-G to the current session, $ w_{ij} $ denotes the weight of edge $ (v_i,v_j) $ in G-G, $ \textbf{q}_1 $ and $ \textbf{W}_1$ are two trainable parameters. $ \odot  $ and $ \parallel $ represent the element-wise product and concatenation operation, respectively.

To make the importance weight comparable across all neighbors connected with $v_i$, we normalize the importance weight of different neighbors by softmax function:
\begin{equation}
  \centering
  b_{ij} = \frac{\exp(b_{ij})}{\sum _{v_{k}\in \mathcal{N} ^{g}_{v_{i}}}\exp(b_{ik})},
\end{equation}
Obviously, the neighbor nodes with higher importance weight should be given more attention.

Then, we calculate the neighbor representation of node $ v_i $ as follows:

\begin{equation}
  \centering
  \textbf{x}_{\mathcal{N} ^{g}_{v_{i}}} = \sum \limits_{ v_j \in  \mathcal{N} ^{g}_{v_{i}}}b_{ij}\textbf{x}_{v_j},
\end{equation}

Finally, the global-level item representations are obtained by aggregating their own representations and their neighbors' representations:
\begin{equation}
  \centering
  \textbf{x}^g_v = Relu(\textbf{W}_2[\textbf{x}_v\parallel \textbf{x}_{\mathcal{N}^g_v}]).
\end{equation}
where $ \textbf{W}_2 $ denotes the trainable parameter and here $ Relu(\cdot) $ is chosen as the activation function.

We denote the pairwise global-level item representations as $\textbf{X}_g$, where each row in $\textbf{X}_g$ represents the local-level item representation after G-GCN learning, i.e. $\textbf{x}^g_{v_i}$.

\subsubsection{H-GCN}

In this section, we first present the definition of H-G. Then we describe the construction process of H-G in detail. Finally, H-GCN is proposed to learn the high-order relations over all the sessions.

\begin{figure}
  \centering
  \includegraphics [scale=0.5]{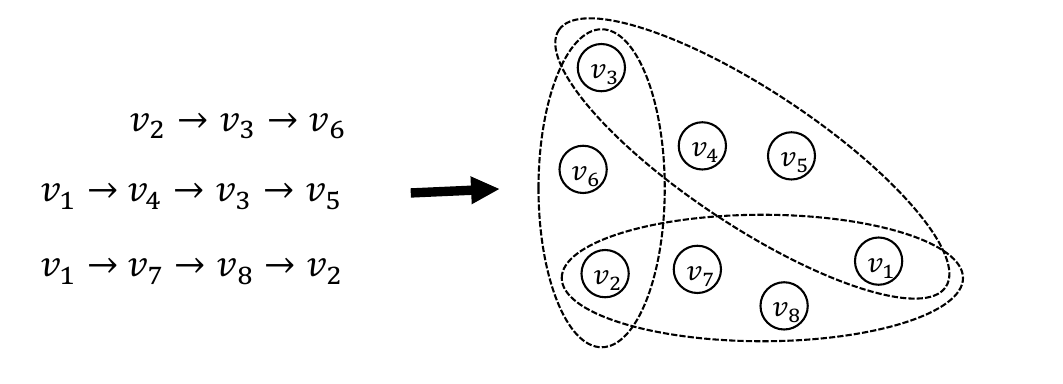}

  \caption  {The construction process of the hyper-level graph.}
  \label{hypergraph}
\end{figure}

Let $ G_h = (\mathcal{V} _h, \mathcal{E} _h)$ denote H-G, where $ \mathcal{V} _h$ and $ \mathcal{E} _h $  denote the set of nodes and the set of hyperedges, respectively. Each hyperedge $ e^h_{ij} \in \mathcal{E}_h $ contains all the items in each session, and  all hyperedges are connected by the shared items in the sessions. The construction process of the H-G is shown in Fig. \ref{hypergraph}.

After the H-G is constructed, we develop the H-GCN to capture high-order item transition patterns. Inspired by S$^2\raisebox{0mm}{-}$DHCN \cite{xia2020self}, we regard the process of H-GCN as a two-stage refinement of the 'node-hyperedge-node' feature transformation. We follow S$^2\raisebox{0mm}{-}$DHCN to assign each hyperedge the same weight with the value of 1 and define H-GCN with row normalization as:

\begin{equation}
  \centering
  \textbf{X}^{t+1}_{h} = \textbf{D}^{-1}\textbf{H}\textbf{W}\textbf{B}^{-1}\textbf{H}^T\textbf{X}^{(t)}_h,
\end{equation}
where $\textbf{H} \in \mathbb{R} ^{(N\times M)}$ indicates the matrix representation of H-G, $\textbf{W}$ is a diagonal matrix which consists of the weight of each hyperedge, $\textbf{D}$ and $\textbf{B}$ are two diagonal matrices which consist of the degree of each vertex and the degree of each hyperedge, respectively.

To get the high-order hyper-level item representations, we extend H-GCN from one layer to multiple layers and average the items representations generated from each layer. Formally, we define this process as follows:

\begin{equation}
  \centering
  \textbf{X}_{h} = \frac{1}{L+1}\sum ^L_{t=0}\textbf{X}^{(t)}_h.
\end{equation}
where $L$ denotes the number of the convolutional layer and $t$ denotes the current H-GCN layer.

We denote each row in $\textbf{X}_{h}$ as the hyper-level item representation after H-GCN learning, i.e., $ \textbf{x}^h_{v_i} $.
\subsection{Session Representation Learning Layer}\label{sess}

Aided by the above three GCNs, we can obtain three different levels of item representations. On this basis, we describe how to learn corresponding session representations.First, We learn the pairwise relation-based item representation by aggregating the representation of items obtained by L-GCN and G-GCN. Specifically, we calculate the above item representation by sum pooling:
\begin{equation}
  \centering
  \textbf{x}^{\prime }_{v_i} = \textbf{x}^g_{v_i} + \textbf{x}^l_{v_i},
\end{equation}

We denote the pairwise relation-based item representations as $\textbf{X}_c$, where each row in $\textbf{X}_c$ represents item representation after the above sum pooling operation, i.e. $\textbf{x}^{\prime }_{v_i}$.

After that, to obtain a more comprehensive session representation, we introduce an attention-based fusion module to adaptively calculate the contribution of different items to the next prediction. First, to incorporate the sequential information into the session representation, we add a learnable position embeddings $ \textbf{p}  \in \mathbb{R}^d  $ into the item representations:
\begin{equation}
  \centering
  \textbf{z}_{v_i} = Tanh(\textbf{W}_3\lfloor \textbf{x}^{\prime }_{v_i} \Vert  \textbf{p}_{t-i+1} \rfloor  + \textbf{b}_3),
\end{equation}
where $ \textbf{W}_3 $ and $ \textbf{b}_3 $ are trainable parameters and here $Tanh(\cdot)$ is chosen as the activation function.

We get the static representation $\textbf{s}^{\prime }$ of the session by averaging item representations in it:
\begin{equation}
  \centering
  \textbf{s}^{\prime } = \frac{1}{t}\sum^t_{i=1}\textbf{x}^{\prime }_{v_i},
\end{equation}

Following SR-GNN \cite{wu2019session}, we calculate the attention coefficients as follows:
\begin{equation}
  \centering
  \beta_i = \textbf{q}^T_2\sigma(\textbf{W}_4\textbf{z}_{v_i} + \textbf{W}_5\textbf{s}^{\prime } + \textbf{b}_4 ),
\end{equation}
where $ \sigma(\cdot ) $ is the sigmod function, and $\textbf{W}_4,\textbf{W}_5,\textbf{q}_2,\textbf{b}_4$ are trainable parameters.

Finally, we obtain the pairwise relation-based session representation $ \textbf{s}_c  $ by calculating the linear combination of the attention coefficient and its corresponding item representation:
\begin{equation}
  \centering
  \textbf{s}_c = \sum^t_{i=1}\beta_i\textbf{x}^{\prime }_{v_i}.
\end{equation}

After obtaining the pairwise relation-based session representation, we obtain the high-order relation-based session representation $ \textbf{s}_h $ by averaging the hyper-level item representations:
\begin{equation}
  \centering
  \textbf{s}_h = \frac{1}{m}\sum^{m}_{i=1}\textbf{x}^h_{v_i}.
\end{equation}

\subsection{Prediction Layer}\label{predic}
In this section, we describe the prediction layer to compute the score of each candidate item based on the learned session representation. Given a session $ s $, we calculate the scores $\hat{\textbf{y}}_{v_i}$ for all the candidate items by multiplying the pairwise relation-based session representation with all item representations:
\begin{equation}
  \centering
  \hat{\textbf{y}}_{v_i} = Softmax(\textbf{s}_c^T\textbf{x}_{v_i}).
\end{equation}
here $Softmax(\cdot)$ is chosen as our activation function.

\subsection{Model Optimization}\label{opti}
In this section, we present the optimization process of our model. Specifically, we difine the learning objective as the cross entropy loss function, which has been extensively used in recommendation systems:
\begin{equation}
  \centering
  \mathcal{L}_r = -\sum^{ n }_{i=1}\textbf{y}_{v_i}log(\hat{\textbf{y}}_{v_i})+(1-\textbf{y}_{v_i})log(1 - \hat{\textbf{y}}_{v_i}).
\end{equation}
where $\textbf{y}$ denotes the one-hot encoding vector of the ground truth item.

\subsection{Enhancing CM-GNN with Contrastive Learning}\label{contras}
In this section, we show how to mine self-supervision signals based on contrastive learning to enhance the training process of the model. In the above subsections, we capture the pairwise item transition patterns over all the sessions through L-GCN and G-GCN, and obtain the pairwise relation-based session representation based on an attention-based fusion module. We also use H-GCN to capture the high-order item transition patterns over all the sessions and obtain the high-order relation-based session representation by averaging the hyper-level item representations. Both the above two levels of session representation only capture a specific item transition patterns in the session, which suffers more from the problem of data sparsity and result in suboptimal recommendation performance. To get a more accurate and comprehensive session representation, we convert the high-order item transition information into the pairwise relation-based session representation. Specifically, we regard the two levels of session representation as two views characterizing different aspects of the same session and innovatively integrate contrastive learning into the training of CM-GNN. CM-GNN improves the recommendation performance by maximizing the mutual information between two different levels of session representation.

Technically, we adopt InfoNCE \cite{oord2018representation}, which can maximize the mutual information between the different levels of session representation, as our learning objective:
\begin{equation}
  \centering
  \mathcal{L}_s = - log\sigma (f_D(\textbf{s}^c_{v_i},\textbf{s}^h_{v_i})) - log\sigma (1  - f_D(\tilde{\textbf{s}}^c_{v_i},\textbf{s}^h_{v_i})),
\end{equation}
where $ f_D(\cdot) : \mathbb{R}^d \times \mathbb{R}^d \longmapsto \mathbb{R} $ is the discriminator function that takes two vectors as the input and then scores the agreement between them. We simply implement the discriminator as the dot product between two representations. Since the pairwise relation-based session representation and the high-order relation-based session representation both model the same session, they can be the ground truth of each other. We corrupt $ \textbf{s}^c_{v_i} $ by both row-wise and column-wise shuffling to create negative examples $ \tilde{\textbf{s}}^c_{v_i} $.

Finally, we unify the recommendation task and the contrastive task and optimize it by means of joint learning. Formally, we define the joint learning objective as follows:
\begin{equation}
  \centering
  \mathcal{L} = \mathcal{L}_r + \beta \mathcal{L}_s.
\end{equation}
where $\beta$ is a hyper-parameter used to control the magnitude of the contrastive task.

\section{Experiments}\label{expe}
In this section, we conduct extensive experiments to validate our model and answer the following four key research questions:

\textbf{RQ\;1}\;Does CM-GNN achieve state-of-the-art performance?

\textbf{RQ\;2}\;Does each component in CM-GNN contribute?

\textbf{RQ\;3}\;How does the number of layers in different GCNs influence the performance of CM-GNN?

\textbf{RQ\;4}\;What is the effectiveness of contrastive learning in CM-GNN?

\textbf{RQ\;5}\;How does hyper-parameter $ \beta  $ in contrastive learning influence the performance of CM-GNN?

\subsection{Datasets}

We evaluate our model on three real-world benchmark datasets, i.e., Diginetica, Tmall, and Nowplaying. The Diginetica\footnote{https://competitions.codalab.org/competitions/11161} dataset is obtained from the CIKM Cup 2016, which consists of typical transaction data. Tmall\footnote{https://tianchi.aliyun.com/dataset/dataDetail?dataId=42} dataset comes from the IJCAI-15 competition, which consists of anonymized users' shopping logs on the Tmall online shopping platform. Nowplaying\footnote{http://dbis-nowplaying.uibk.ac.at/nowplaying} is obtained from \cite{zangerle2014nowplaying}, which consists of the music listening behavior of users. For a fair comparison, we follow the experiment environment in \cite{xia2020self} to preprocess the three benchmark datasets. Specifically, we filter out all the sessions containing only one item and items appearing less than five times for all the datasets. We regard the latest data (such as the sessions of last week) as test data and the remaining data as the training set. Moreover, we augment and label both the training dataset and the test dataset by employing a sequence splitting approach for all the datasets, then we generate multiple labeled sequences with the corresponding labels, i.e.,$$ ([v_{s,1}],v_{s,2}),([v_{s,1},v_{s,2}],v_{s,3}),...,([v_{s,1},v_{s,2},v_{s,m-1}],v_{s,m}).$$
\noindent We summary the statistics of datasets after preprocessing in Table \ref{dataset}.

\begin{table}
  \caption{Dataset Statistics}
  \centering
  \scalebox{1.2}{
  \begin{tabular}{cccc}
  \toprule
  Dataset & Tmall  & Diginetica & Nowplaying \\
  \hline \hline
  training sessions  & 351268 & 719470 & 825304 \\
  test sessions  & 25898 & 60858 & 89824 \\
  \# of items & 30968 & 40728 & 43097 \\
  average lengths& 6.69 & 5.12 & 7.42 \\
  \bottomrule
  \label{dataset}
  \end{tabular}}
  \vspace{-7mm}
\end{table}

\subsection{Baseline Methods}
We compare CM-GNN with eleven strong and commonly used methods, which are presented as follows:

\begin{itemize}
  \item[$\bullet$] \textbf{POP}: POP recommends top-K items based on their popularity.
  \item[$\bullet$] \textbf{Item-KNN}\cite{sarwar2001item}: Item-KNN computes the cosine similarity between items of the current session and other sessions and recommends the most similar top-K items.
  \item[$\bullet$] \textbf{FPMC}\cite{rendle2010factorizing}: FPMC captures users' short-term preferences by combining the matrix factorization and the first-order Markov chain.
  \item[$\bullet$] \textbf{GRU4REC}\cite{hidasi2015session}: GRU4REC employs Gated Recurrent Unit (GRU) to model user behavior sequences.
  \item[$\bullet$] \textbf{NARM}\cite{li2017neural}: NARM is an RNN-based method, which adopts the attention mechanism to model sequential patterns.
  \item[$\bullet$] \textbf{STAMP}\cite{liu2018stamp}: STAMP utilizes attention layers to replace all RNN encoders in the previous work and resorts to the last item as the user's short-term interest in the current session to make the prediction.
  \item[$\bullet$] \textbf{CSRM}\cite{wang2019collaborative}: CSRM utilizes a parallel memory module to capture the sequential patterns of the latest m sessions.
  \item[$\bullet$] \textbf{SR-GNN}\cite{wu2019session}: SR-GNN applies a Gated Graph Neural Network (GGNN) to learn item embeddings and obtains the session embeddings by applying a soft-attention mechanism.
  \item[$\bullet$] \textbf{FGNN}\cite{qiu2019rethinking}: FGNN extends NARM by considering the inherent order of the item transition patterns.
  \item[$\bullet$] \textbf{GCE-GNN}\cite{wang2020global}: GCE-GNN applies different levels of graph convolutional networks (GCNs) on both session-level graph and global-level graph respectively and captures comprehensive pairwise item transition patterns.
  \item[$\bullet$] $\textbf{S}^2\raisebox{0mm}{-}\textbf{DHCN}$\cite{xia2020self}: S$^2\raisebox{0mm}{-}$DHCN constructs a hypergraph to capture the high-order relations and constructs a linegraph for contrastive learning to enhance network training.

\end{itemize}

\subsection{Evaluation Metrics}
Following the previous work \cite{wu2019session}, \cite{xia2020self}, \cite{wang2020global}, we adopt two relevancy-based metrics P@K(Precision) and MRR@K(Mean Reciprocal Rank) to evaluate our approach. In the top-K items, P@K indicates the proportion of correctly recommended items. MRR@K represents the average of reciprocal ranks of the correctly recommended items in the top-k ranking list.
In the paper, we adopt P@K and MRR@K with K = 10, 20 to evaluate all compared methods.

\subsection{Parameter Setup}
For CM-GNN, we split the random 10\% subset of the training set into the validation set and keep the hyper-parameters of each model consistent for a fair comparison. We set the dimension of latent vectors to 100 for all the datasets. Besides, we set mini-batch to 256 and we use Gaussian distribution $\mathcal{N}(0,0.1^2)$ to initialize all parameters in our model. We use the Adam optimizer to optimize these parameters, where the initial learning rate is set to 0.001 and decay by 0.1 after every 3 epochs. The L2 penalty is set to $10^{-5}$ and the number of neighbors and the maximum distance of adjacent items $ \varepsilon  $ are set to 12 and 3, respectively. Here, we follow their best parameter setups reported in the original papers and present their best results for all the baseline methods.

\begin{table*}[]
  \caption{The overall performances based on three datasets. The best is bold and the second is underline.}
  \begin{tabular}{lcccc|cccc|cccc}
  \toprule
  Datasets & \multicolumn{4}{c|}{Tmall}       & \multicolumn{4}{c|}{Diginetica}  & \multicolumn{4}{c}{Nowplaying}  \\ \hline
  Methods  & P@10  & P@20  & MRR@10 & MRR@20 & P@10  & P@20  & MRR@10 & MRR@20 & P@10  & P@20  & MRR@10 & MRR@20 \\ \hline\hline
  POP      & 1.67  & 2.00  & 0.88   & 0.90   & 0.76  & 1.18  & 0.26   & 0.28   & 1.86  & 2.28  & 0.83   & 0.86   \\
  Item-KNN & 6.65  & 9.15  & 3.11   & 3.31   & 25.07 & 35.75 & 10.77  & 11.57  & 10.96 & 15.94 & 4.55   & 4.91   \\
  FPMC     & 13.10 & 16.06 & 7.12   & 7.32   & 15.43 & 22.14 & 6.20   & 6.66   & 5.28  & 7.36  & 2.68   & 2.82   \\
  GRU4Rec  & 9.47  & 10.93 & 5.78   & 5.89   & 17.93 & 30.79 & 7.73   & 8.22   & 6.74  & 7.92  & 4.40   & 4.48   \\
  NARM     & 19.17 & 23.30 & 10.42  & 10.70  & 35.44 & 48.32 & 15.13  & 16.00  & 13.6  & 18.59 & 6.62   & 6.93   \\
  STAMP    & 22.63 & 26.47 & 13.12  & 13.36  & 33.98 & 46.62 & 14.26  & 15.13  & 13.22 & 17.66 & 6.57   & 6.88   \\
  CSRM     & 24.54 & 29.46 & 13.62  & 13.96  & 36.59 & 50.55 & 15.41  & 16.38  & 13.20 & 18.14 & 6.08   & 6.42   \\
  SR-GNN   & 23.41 & 27.57 & 13.45  & 13.72  & 38.42 & 51.26 & 16.89  & 17.78  & 14.17 & 18.87 & 7.15   & 7.47   \\
  FGNN     & 20.67 & 25.24 & 10.07  & 10.39  & 37.72 & 50.58 & 15.95  & 16.84  & 13.89 & 18.78 & 6.8    & 7.15   \\
  GCE-GNN  & \underline{28.01} & \underline{33.42} & \underline{15.08}  & \underline{15.42}  & \underline{41.16} & \underline{54.22} & \underline{18.15}  & \underline{19.04}  & 16.94 & 22.37 & \underline{8.03}   & \underline{8.40}   \\
  S$^2\raisebox{0mm}{-}$DHCN     & 26.22 & 31.42 & 14.60  & 15.05  & 40.21 & 53.66 & 17.59  & 18.51  & \underline{17.35} & \textbf{23.50} & 7.87   & 8.18   \\ \hline\hline
  \textbf{CM-GNN}   & \textbf{29.72} & \textbf{34.69} & \textbf{16.06}  & \textbf{16.41}  & \textbf{41.56} & \textbf{54.67} & \textbf{18.28}  & \textbf{19.19}  & \textbf{17.50} &\underline{23.30} & \textbf{8.05}   & \textbf{8.44}   \\ \bottomrule
  \end{tabular}
  \label{performance}
\end{table*}

\begin{table*}[]
  \caption{The ablation study.}
  \begin{tabular}{lcccc|cccc|cccc}
  \toprule
  Dataset         & \multicolumn{4}{c|}{Tmall}                                         & \multicolumn{4}{c|}{Diginetica}                                    & \multicolumn{4}{c}{Nowplaying}                                  \\ \hline
  Measures        & P@10           & MRR@10         & P@20           & MRR@20         & P@10           & MRR@10         & P@20           & MRR@20         & P@10           & MRR@10        & P@20           & MRR@20        \\ \hline\hline
  CG-L        & 12.10          & 6.64           & 14.56          & 6.81           & 35.50          & 14.42          & 48.71          & 15.34          & 15.42          & 7.47          & 20.69          & 7.83          \\
  CG-G        & 25.94          & 13.86          & 30.41          & 14.17          & 41.53          & 18.14          & 54.53          & 19.05          & 17.49          & 7.36          & 22.89          & 7.76          \\
  CG-H        & 28.85          & 15.40          & 33.85          & 15.76          & 41.55          & 18.21          & 54.53          & 19.11          & 17.35          & 8.12          & 22.79          & 8.49          \\
  CG-OL        & 28.55          & 14.78          & 32.74          & 15.02          & 37.25          & 15.77          & 51.34          & 16.71          & 16.47          & 7.09          & 22.10          & 7.35          \\
  CG-OG        & 18.29          & 9.82          & 20.54          & 10.71          & 30.80          & 12.02          & 45.58         & 13.42          & 14.47          & 6.46          & 18.39          & 6.64          \\
  CG-OH        & 26.03          & 12.44          & 31.27          & 13.93          & 37.88          & 16.01          & 52.18          & 16.84          & 15.63          &  6.72         & 21.34          & 7.02          \\
  CG-A        & 24.49          & 13.94          & 28.98          & 14.25          & 36.34          & 15.90          & 48.48          & 16.73          & 13.32          & 5.71          & 18.83          & 6.09          \\
  CG-S        & 28.54          & 15.14          & 33.70          & 15.50          & 41.39          & 18.20          & 54.55          & 19.11          & 17.36          & \textbf{8.13}          & 22.93          & \textbf{8.52}          \\
  \textbf{CM-GNN} & \textbf{29.72} & \textbf{16.07} & \textbf{34.69} & \textbf{16.41} & \textbf{41.56} & \textbf{18.28} & \textbf{54.69} & \textbf{19.19} & \textbf{17.83} & 8.05 & \textbf{23.30} & 8.44 \\ \bottomrule
  \end{tabular}
  \label{variants}
\end{table*}

\subsection{Recommendation Performance (RQ 1)}

To evaluate the performance of CM-GNN, we report the comparison results on three widely used benchmark datasets in Table 2, where we highlight the best result of each column in boldface. As can be observed, our proposed CM-GNN outperforms all the baseline methods on all datasets.

Traditional methods include POP, FPMC, and Item-KNN. POP only recommends top-N frequent items according to popularity, showing the worst performance. FPMC utilizes first-order Markov chain and matrix factorization to recommend the next item by taking into account the simple contextual information of the session, which achieves better performance than POP. Among the traditional methods, Item-KNN achieves the best results, which recommends the top-K items by computing the cosine similarity between items of the current session and other sessions. However, Item-KNN fails to capture the sequential patterns within the session and still does not achieve the desired performance.

Compared with the conventional methods, neural network-based methods consider the sequential dependency in the sessions and achieve decent performance for SBR. Methods based on deep recurrent neural structure are the earliest deep learning methods for modeling user preferences. Specifically, GRU4REC is the first work to adapt Recurrent Neural Networks (RNNs), which employs Gated Recurrent Unit (GRU) to model user behavior sequences. Compared with GRU4Rec, NARM and STAMP introduce the attention mechanism into RNN and improve the performance by a large margin. The superior performance of NARM and STAMP indicates that the attention mechanism has played an important role in sequence modeling. However, NARM and STAMP are modeled only based on the explicit interactions of user behavior, which neglects the implicit connection of the items in sessions. By incorporating collaborative information from others into the current session, CSRM enrichs the session representation and achieves decent performance. However, CSRM often encodes irrelevant information from other sessions into the current session and suffers from noise impact. To overcome the above issues, subsequent studies begin to model session-based data as graphs and employ Graph Neural Networks (GNNs) to capture implicit connections among items in sessions. SR-GNN is the first work to employ GNN for sequential patterns capturing in sessions, which captures pairwise relations among items based on the Graph Gating Neural Network (GGNN). Compared with SR-GNN, FGNN takes into account the specific item transition patterns within the session for better performance. However, SR-GNN and FGNN only capture the sequential patterns in the current session and do not take advantage of item transition information in other sessions. GCE-GNN captures pairwise item transition information in all the sessions by constructing session-based data into a session-level graph and global-level graph, respectively. However, GCE-GNN only captures pairwise relations among items in sessions which neglects the complex and high-order item transition information. In recent work, S$^2\raisebox{0mm}{-}$DHCN has achieved decent recommended performance by modeling high-order relations among items in sessions. However, S$^2\raisebox{0mm}{-}$DHCN fails to explicitly capture pairwise item transition information among items. Besides, since sessions are generally short sequences formed by the interactions of anonymous users, modeling session-based data directly as a hypergraph can suffer from data sparsity problem.

Compared with the above baselines, our proposed CM-GNN shows overwhelming superiority on all evaluation metrics over all the datasets. On average, CM-GNN outperforms the best result by 3.7\% on Tmall, 0.8\% on Deginetica, 3.3\% on Nowplaying. CM-GNN shows significant recommendation advantages over most methods, which are benefit from the consideration of global item transition information. By capturing the item transition patterns over all the sessions, CM-GNN generates more comprehensive and accurate session representations. Compared with GCE-GNN, CM-GNN still achieves decent performance improvements on the Tmall and Nowplaying datasets, which attributes to the ability to model high-order relations based on hyper-level graph. In contrast, CM-GNN's performance improvement on the Diginetica dataset was not significant, one possible explanation is that the average session length of Diginetica is relatively smaller than the other two datasets, which makes the modeling of high-order relations does not capture more effective session features.

\subsection{Ablation Study}
The overwhelming superiority of CM-GNN presented in the last section can be seen as the result of joint effect of different components. To investigate the contributions of each component in CM-GNN, we conduct extensive experiments on variants of CM-GNN for all the datasets.

\subsubsection{Impact of each component in CM-GNN (RQ2)}
In this subsection, we conduct extensive experiments on three datasets to evaluate the effectiveness of local-level graph convolutional network (L-GCN) module, global-level graph convolutional network (G-GCN) module, hyper-level graph convolutional network (H-GCN) module, attention-based fusion module and contrastive learning module. Technically, we develop eight variants of CM-GNN: CG-L, CG-G, CG-H, CG-OL, CG-OG, CG-OH, CG-A, and CG-S. Specifically, CG-L represents the version without L-GCN module, CG-G represents the version without G-GCN module, CG-H represents the version without H-GCN module, CG-OL represents the version that we remove G-GCN module and H-GCN module and only remain L-GCN module, CG-OG represents the version that we remove L-GCN module and H-GCN module and only remain G-GCN module, CG-OH represents the version that we remove L-GCN module and G-GCN module and only remain H-GCN module,  CG-A represents the version that we remove the attention-based fusion module and replace it with averaging item representations as the representation of each session and CG-S represents the version that we replace the InfoNCE loss function with MSE loss function. We report the results of CM-GNN and these five variants above in Table \ref{variants}.

As can be observed from Table \ref{variants}, CM-GNN's performance on both the Tmall and Diginetica datasets consistently outperforms the five variants, which demonstrates that each component contributes to the CM-GNN recommendation process. Overall, L-GCN module is one of the most critical parts of CM-GNN, which serves for the capture of the item transition patterns within the current session. When removing L-GCN module, we can observe a remarkable performance drop on all the four metrics on both the Tmall dataset and Diginetica dataset. Correspondingly, it can be observed that all the metrics on the Nowplaying dataset have a significant drop when the attention-based fusion module is replaced with the average pooling module, which demonstrates that the attention-based fusion module avoids bringing up a bias of unrelated items when propagating information from all items within all the sessions. For the Tmall and Diginetica datasets, G-GCN module and the attention-based fusion module are the modules that play the second key role in CM-GNN, while for the Nowplaying dataset, the second most important modules of CM-GNN are L-GCN module and G-GCN module. We can conclude that the capture of global-level item transition patterns helps the model make more accurate predictions and learning different item importance across sessions is better than directly averaging representations of contained items for learning session representations in SBR. Moreover, Table \ref{variants} shows that all the metrics on the Tmall dataset and the Diginetica dataset have a significant drop when either the H-GCN module is removed or the InfoNCE loss function is replaced with MSE loss, which indicates that the application of H-GCN module and contrastive learning help the model learn more high-order information. Relatively speaking, although H-GCN module and the contrastive learning module also contribute to the P@K metric on the Nowplaying dataset, they do not exhibit competitive performance on the MRR@K metric. One possible explanation is that due to the relatively small average session length of Nowplaying, the pairwise item transition patterns and the high-order item transition patterns captured by CM-GNN may be very similar, which results in the model collapse when training based on contrastive learning.

\subsubsection{Impact of Model Depth (RQ3)}
We evaluate the influence of the number of layers in different GCNs on CM-GNN recommendation performance. Specifically, we perform all the experiments on our RTX 2080 Ti GPU, and to avoid suffering from the out-of-memory problem, we fix the number of layers of L-GCN to 1. Moreover, we range the number of layers of G-GCN in \{1,2\} and we range the number of layers of H-GCN in \{1,2,3\}. Technically, We develop four variant versions of CM-GNN: G-1-H-1, G-2-H-1, G-2-H-2, and G-2-H-3 to explore the influence of the number of layers of different levels in GCNs on performance. In G-1-H-1, we set the number of layers of G-GCN and H-GCN to 1 and 1, respectively. In G-2-H-1, we set the number of layers of G-GCN and H-GCN to 2 and 1, respectively. In G-2-H-2, we set the number of layers of G-GCN and H-GCN to 2 and 2, respectively. In G-2-H-3, we set the number of layers of G-GCN and H-GCN to 2 and 3, respectively. We show the results of four variants of CM-GNN in Table \ref{depth}.

\begin{table}[]
  \caption{The comparison performances of different layers in GCNs module.}
  \scalebox{0.9}{
  \begin{tabular}{lcc|cc|cc}
  \toprule
  Datasets         & \multicolumn{2}{c|}{Tmall}       & \multicolumn{2}{c|}{Diginetica}  & \multicolumn{2}{c}{Nowplaying} \\ \hline
  Measures         & P@20           & MRR@20         & P@20           & MRR@20         & P@20           & MRR@20        \\  \hline\hline
  G-1-H-1          & 28.92          & 14.08          & 54.30          & 19.14          & 23.17          & 8.30          \\
  \textbf{G-2-H-1} & \textbf{34.56} & \textbf{16.41} & \textbf{54.66} & \textbf{19.19} & \textbf{23.24} & 8.39          \\
  G-2-H-2          & 34.53          & 16.33          & 54.61          & 19.17          & 23.12          & 8.41          \\
  G-2-H-3          & 34.02          & 16.09          & 54.56          & 19.15          & 23.13          & \textbf{8.44} \\ \bottomrule
  \end{tabular}}
  \label{depth}
\end{table}

\begin{figure}
  \centering
  \includegraphics [scale=0.53]{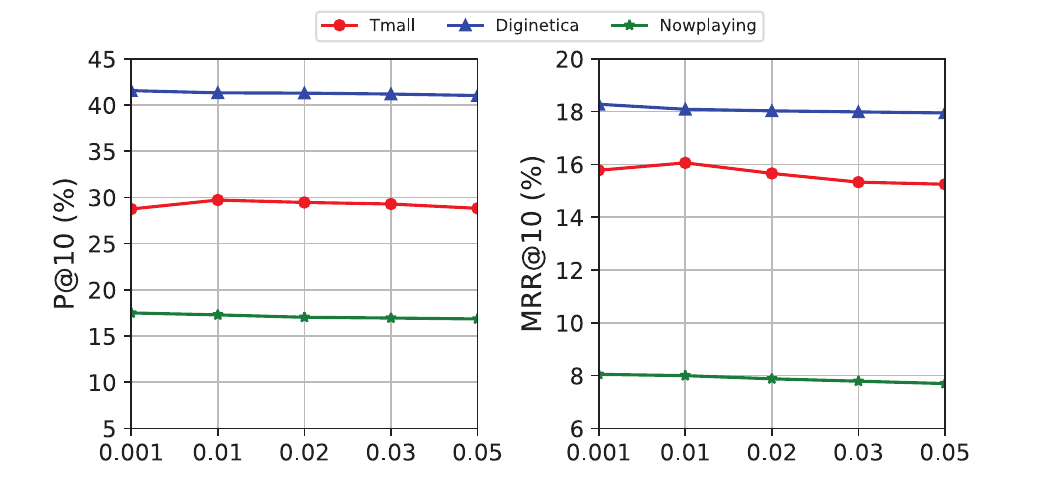}
  \includegraphics [scale=0.53]{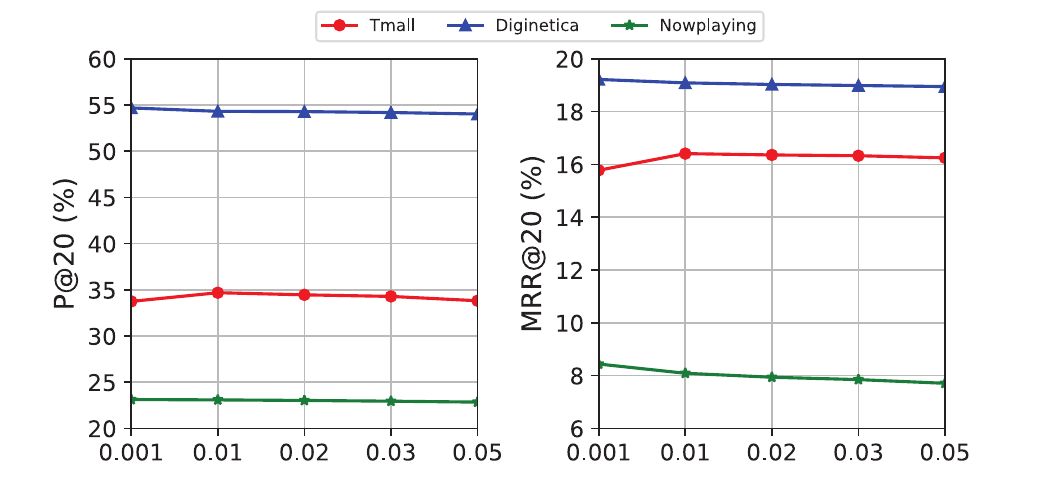}
  \caption  {The influence of the magnitude of contrastive learning.}
  \label{sensitivity}
\end{figure}

From the table \ref{depth},  CM-GNN with G-GCN building by two layers performs better one layer, which indicates that high-level exploring might capture more effective information from global-level graph. Furthermore, we fix the number of layers of the G-GCN to 2 and range the number of layers of H-GCN in \{1, 2, 3\} for comparison. It can be observed that H-GCN with 1 layer performs the best on both the Tmall dataset and Diginetica dataset. As the number of layers of the H-GCN increases, the overall performance tends to drop, which indicates that higher-level exploring might also introduce noise. However, The H-GCN with more layers achieves better performance on the Nowplaying dataset, one possible explanation is that due to the relatively small average session length of Nowplaying, multi-layer H-GCN can capture more item transition patterns.

\begin{table}[]
  \caption{The effectiveness of contrastive learing in CM-GNN.}
  \scalebox{0.9}{
  \begin{tabular}{lcc|cc|cc}
  \toprule
  Datasets         & \multicolumn{2}{c|}{Tmall}       & \multicolumn{2}{c|}{Diginetica}  & \multicolumn{2}{c}{Nowplaying} \\ \hline
  Measures         & P@20           & MRR@20         & P@20           & MRR@20         & P@20           & MRR@20        \\  \hline\hline
  CG-POOL          & 31.21          & 14.47          & 50.72          & 16.22          & 21.71          & 7.87          \\
  \textbf{CM-GNN} & \textbf{34.69} & \textbf{16.41} & \textbf{54.67} & \textbf{19.19} & \textbf{23.30} & \textbf{8.44}          \\
  \bottomrule
  \end{tabular}}
  \label{effectiveness}
\end{table}

\subsubsection{Effectiveness of Contrastive Learning (RQ4)}
Existing methods typically utilize the item prediction loss to learn the entire model. Since the length of session-based data is short, learning parameters of the model with a single optimization objective usually suffers from data sparsity. To this end, we borrow the idea of contrastive learning for alleviating the data sparsity problem and improving recommendation performance. To study the effectiveness of using contrastive learning, we design a variant version of CM-GNN named CG-POOL for comparison on all datasets with the complete CM-GNN model. Specifically, CG-POOL removes contrastive learning in CM-GNN and directly fuses local-level item representations, global-level item representations, and hyper-level item representations for learning final session representation by the attention-based fusion module.

As can be observed from table \ref{effectiveness}, CM-GNN's performance consistently outperforms the variant version CG-POOL across different datasets, which is credited to the capability of contrastive learning on learning the extra self-supervision signals and alleviating the data sparsity problem. Compared with CG-POOL, CM-GNN use contrastive learning to maximize the mutual information between different levels of session representation. And in the above process, different levels of item transition information (e.g. pairwise and high-order item transition information) can learn from each other, while CG-POOL integrates high-level information with pairwise information without differentiating and thus drops structural information in different levels of session representation.

\subsection{Parameter Sensitivity Analysis (RQ5)}
We introduce a hyper-parameter to CM-GNN to control the magnitude of contrastive learning, i.e.,  $ \beta $. To investigate the influence of it, we report the performance with a set of representative $ \beta $ values in \{ 0.001, 0.01, 0.02, 0.03, 0.05 \}. As can be seen in Fig. \ref{sensitivity}, our model performs the best on Diginetica and Nowplaying when $ \beta $ is 0.001, while for Tmall, setting $ \beta $ to 0.01 achieves the best performance. Besides, it can be observed that as $ \beta $ increases, the performance of CM-GNN drops. One possible reason is that there is a gradient conflict between the recommendation task and the contrastive task, which means that when we use the contrastive task to train CM-GNN, we should choose an appropriate $ \beta $ value.

\section{Conclusion}\label{conclu}
In this paper, we propose a novel contrastive multi-level graph neural network (CM-GNN) for SBR. Specifically, we design a local-level graph convolutional network (L-GCN) module and a global-level graph convolutional network (G-GCN) module to capture pairwise relations of the current session and all the sessions, respectively. CM-GNN introduce an attention-based fusion module to learn the pairwise relation-based session representation by fusing the item representations generated by the above modules. Moreover, since existing methods fail to capture the high-order item transition information, CM-GNN designs a hyper-level graph convolutional network (H-GCN) module to capture the high-order item transition patterns. CM-GNN learns the high-order relation-based session representation by averaging the item representations outputted by H-GCN module. Finally, to convert the high-order item transition information into the pairwise relation-based session representation, we apply contrastive learning between different levels of session representation to enhance the training process of the network. In this paper, we have carried out a large number of experiments on CM-GNN, and comprehensive studies show that our CM-GNN outperforms eleven baselines over three benchmark datasets consistently.

In the future, we plan to study the model collapse problem caused by contrastive learning and consider applying a divergence constraint during the training process of CM-GNN to alleviate this problem.

\section*{Acknowledgment}
This work was supported in part by National Natural Science Foundation of China (Nos. 61976127, 61702491).

\bibliographystyle{IEEEtran}
\bibliography{sbr-tmm}

\end{document}